# MOLECULAR DYNAMICS SIMULATIONS OF THE AFFINITY OF CHITIN AND CHITOSAN FOR COLLAGEN: THE EFFECT OF pH AND THE PRESENCE OF SODIUM AND CALCIUM CATIONS

**Maciej Przybyłek[1,a,\*], Piotr Bełdowski[2,b]**

[1] – Department of Physical Chemistry, Pharmacy Faculty, Collegium Medicum of Bydgoszcz, Nicolaus Copernicus University in Toruń, Kurpińskiego 5 Str., 85–950 Bydgoszcz, Poland
[a] – ORCID: 0000–0002–3399–6129
[2] – Faculty of Chemical Technology and Engineering, Institute of Mathematics & Physics, Bydgoszcz University of Science & Technology, prof. S. Kaliskiego 7 Ave., 85–796 Bydgoszcz, Poland
[b] – ORCID:0000–0002–7505–6063
\*corresponding author: m.przybylek@cm.umk.pl

**Abstract**

*Chitosan and chitin are promising biopolymers used in many areas including biomedical applications, such as tissue engineering and viscosupplementation. Chitosan shares similar properties with hyaluronan, a natural component of synovial fluid, making it a good candidate for joint disease treatment. The structural and energetic consequences of intermolecular interactions are crucial for understanding the biolubrication phenomenon and other important biomedical features. However, the properties of biopolymers, including their complexation abilities, are influenced by the nature of the aqueous medium with which they interact. In this study, we employed molecular dynamics simulations to describe the effect of pH and the presence of sodium and calcium cations on the stability of molecular complexes formed by collagen type II with chitin and chitosan oligosaccharides. Based on Gibbs free energy of binding, all considered complexes are thermodynamically stable over the entire pH range. The affinity between chitosan oligosaccharide and collagen is highly influenced by pH, while oligomeric chitin shows no pH-dependent effect on the stability of molecular assemblies with collagen. On the other hand, the presence of sodium and calcium cations has a negligible effect on the affinity of chitin and chitosan for collagen.*

# 1. Introduction

Synthesis of biopolymer composites and blends is a popular strategy to obtain materials with tailored physicochemical and mechanical properties. Furthermore, various biopolymers commonly co-exist in the human body, including peptides and hyaluronic acid in the extracellular matrix [1, 2]. As this example shows, protein-polysaccharide interactions are particularly interesting from medical and biological viewpoints. Molecular complexes formed by these systems seem to be crucial in describing the biolubrication phenomenon that involves cartilage and synovial fluid [3–6]. Of note, hyaluronate formulations administered via intra-articular injections are commonly used for viscosupplementation [7, 8]. Chitosan is another biopolymer that has been considered for this purpose [9, 10]. This popular macromolecule can be obtained from chitin via chemical [11–14] or enzymatic [15–17] deacetylation. The chemical method, which is carried out in the presence of strong bases, is more common; however, it has some disadvantages mainly associated with polymer chain hydrolysis [13].

Both chitosan and chitin are relatively inexpensive and 'green' materials [18, 19]. These biocompatible polymers are characterised by unique properties and features including antioxidant and antibacterial activities [20–22] and the ability to form hydrogels, films, membranes, nanofibres, sponges, and scaffolds [23]. Collagen is an extracellular matrix molecule that plays an important role in regeneration, maintaining the integrity and optimal mechanical properties of various tissues such as skin, blood vessels, cartilage, and bones [24–26]. Biomaterials containing chitosan, chitin, and collagen have been utilised in tissue engineering [27–30], cosmetics [31–33], and drug delivery [34, 35]. These two glycosaminoglycan analogues exhibit a significant affinity for collagen, which has been proven experimentally [36] and theoretically [4, 5]. Notably, chitosan has often been used to prepare collagen scaffolds with excellent mechanical properties [37–40].

Molecular dynamics (MD) is a very popular method applied as part of biomaterial design [41]. The main advantages offered by this approach are efficiency and sufficiently high accuracy when describing protein-polysaccharide complexes or analogous systems [42, 43]. Although collagen blends with hyaluronic acid, chitin, and chitosan are very popular, as evidenced by the above-mentioned examples, there are not many MD studies on these systems. In our previous papers, we described the two important structural factors, namely the degree of deacetylation of chitosan [4, 5] and the proline/hydroxyproline (Pro/Hyp) ratio in collagen [5, 6] using MD simulations. In the present study, we evaluated the effect of pH and the presence of sodium ($Na^+$) and calcium ($Ca^{2+}$) cations on the affinity of chitin and chitosan oligosaccharides for collagen by using a similar computational approach. While the utilisation of low-molecular-weight oligomeric species is associated with methodological limitations of the applied approach, it is important to highlight that chitooligosaccharides are promising materials for biomedical and pharmaceutical applications [44–47].

# 2. Methods
## 2.1. Computational Details

The collagen type II sequence (Pro–Hyp–Gly)$_3$–Arg–Ala–Gly–Glu–Pro–Gly–Leu–Gln–Gly–Pro–Ala–Gly–(Pro–Hyp–Gly)$_3$ was obtained from the RCSB PDB database [48]. The selected structure contains essential residues (12 amino acid moieties) that form a common motif characteristic of human collagen type II. To improve the stability of the triple-helical system, the structures were modified by adding terminal triplets, specifically (Pro–Hyp–Gly)$_3$. In this work, trans–4–hydroxy–L–proline was applied as the Hyp residue. The structure of chitin with a molecular mass of 1643.6 Da was downloaded from the PubChem database [49] and modified to ~3.3 kDa. A chitosan molecule with a degree of deacetylation of 100% was prepared from the PubChem chitin structure.

The accurate three-dimensional (3D) structures of the chitooligosaccharides were determined using a geometry optimisation procedure. For the docking step, the default VINA approach [50] was applied. The initial charges of the peptide were calculated by utilising the AMBER14 force field [51]. In the case of chitosan/chitin molecules, the GLYCAM06 force field adapted for the AutoDock optimisation protocol (scoring function-based procedure) [52] was used. The applied simulation boxes with the considered molecular assemblies contained additional solvent molecules (water). The total neutral charge of the systems was maintained by chloride counterions ($Cl^-$). The most energetically favoured structures were selected from the considered 50 runs. The applied free binding energy threshold was 42 kJ/mol. This procedure led to the selection of 150 complexes of collagen type II with chitin or chitosan oligosaccharides. All the above-mentioned simulations were performed with an aid of YASARA software [53].

Because all calculations were carried out in the aqueous media, the reasonable acid-base equilibrium microstates were adjusted [54]. All manipulations used to determine the considered effects were performed using standard protocols available in YASARA. The pH range from 1 to 14 was selected to understand the influence of pH on intermolecular interactions. The effect of the presence of $Ca^{2+}$ and $Na^+$ was determined by using a mass concentration of 0.9%. After reaching the steepest gradient in the annealing simulations, MD calculations were performed with a different force field for each molecule in the considered systems: AMBER14 (collagen), GLYCAM06 (chitin/chitosan), and TIP3P (water). There was no threshold considered in the case of the Particle Mesh Ewald protocol [55], which was applied at the electrostatic contact detection step [56], while in the case of van der Waals forces, the 10–Å threshold was included (the default AMBER criterion). The standard parameters were utilised to integrate motion equations, namely multiple time steps of 1.25 and 2.5 fs for bonded interactions in the case of non-bonded forces and isothermal-isobaric conditions (temperature = 310 K, pressure = 1 atm) [53]. As a result of root mean square deviation/displacement (RMSD) analysis, 100 points of free energy of binding, the number of hydrogen bonds (H–bonds), ionic and hydrophobic (HP) contacts, as well as the number of water bridges in the 0.9–1 ns range from MD simulations (0.01–ns sampling) were selected for additional calculations, including determination of the binding free energy.

### 2.2. Determination of the Binding Free Energy

The binding Gibbs free energy ($\Delta G_b$), which is defined by Equation 1, was determined with the aid of a single trajectory method using the YASARA Structure and Molecular Mechanics/Poisson–Boltzmann Surface Area modules. The symbols $\Delta G_{complex}$, $\Delta G_{chitin/chitosan}$, and $\Delta G_{collagen}$ denote Gibbs free energy values computed for the complex, chitin/chitosan structures, and collagen, respectively.

$$\Delta G_b = \Delta G_{complex} - (\Delta G_{chitin/chitosan} + \Delta G_{collagen}) \qquad (1)$$

Solvation energy and electrostatic effects were considered by using the Adaptive Poisson–Boltzmann Solver (APBS) and the AMBER14 force field [57, 58].

## 3. Results and Discussion

In the first stage of the study, we examined the effect of pH on the affinity of chitooligosaccharides for collagen type II. Figures 1 and 2 present examples of optimised collagen complexes with chitosan and chitin oligosaccharides, respectively, at pH 1, 7, and 14.

**pH=1**

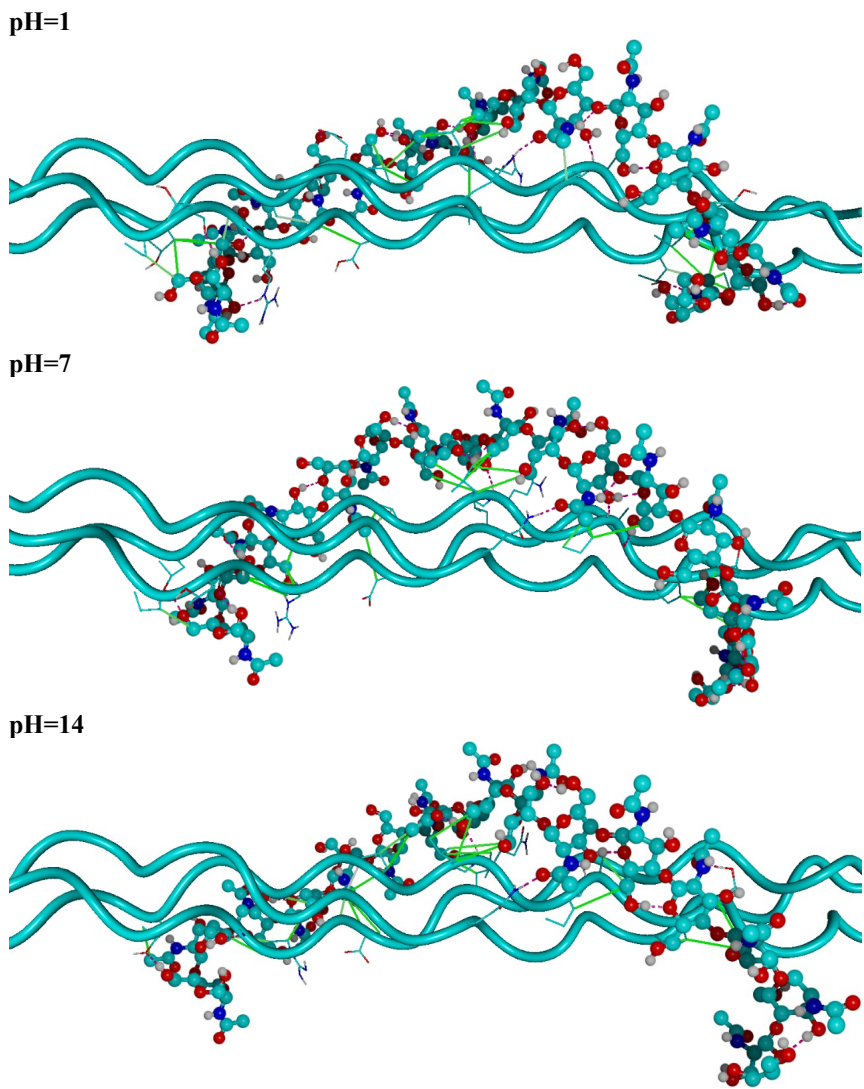

**pH=7**

**pH=14**

**Figure 1.** Graphical representation of the selected most stable collagen type II molecular assemblies with chitin oligosaccharides. The pink dotted lines indicate hydrogen bonds, while the solid green lines denote hydrophobic contacts. Ionic contacts do not occur due to the non-polyelectrolytic nature of chitin.

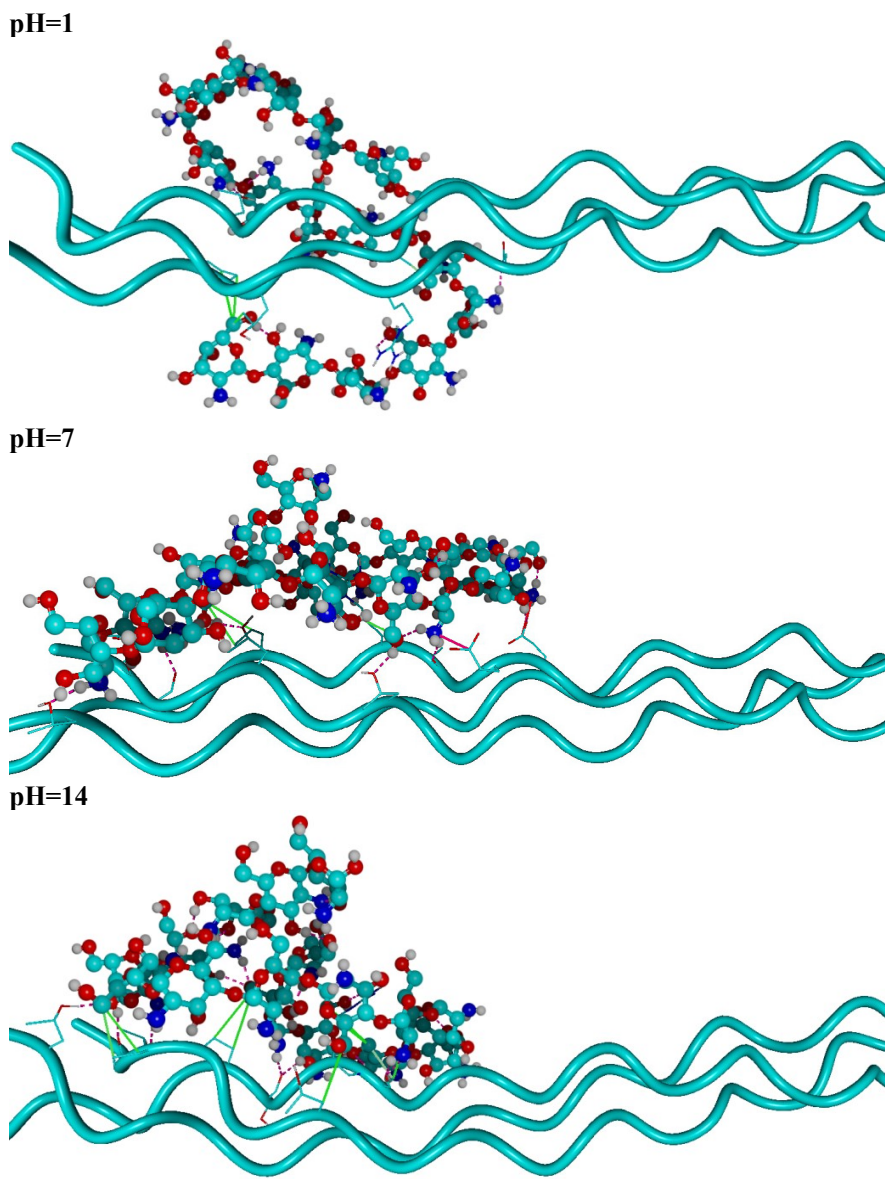

**Figure 2.** Graphical representation of the selected most stable collagen type II molecular assemblies with chitosan oligosaccharides. The pink dotted lines indicate hydrogen bonds, the solid pink lines indicate ionic interactions, and the solid green lines represent hydrophobic contacts.

It should be noted that collagen is relatively stable in a wide range of pH [59]. However, its denaturation temperature is higher in alkaline conditions than in acidic ones [59, 60]. On the other hand, chitin undergoes deacetylation in alkaline aqueous media. Although we analysed the full pH range, it should be noted that in the case of a highly alkaline environment, chitin is less stable than chitosan. Nevertheless, chitin deacetylation is a very slow process at room temperature [61, 62] and thus the reaction is often carried out at elevated temperatures [11]. Therefore, it is reasonable to investigate potential chitin-collagen complexes at the temperature used in the MD simulations (i.e., 37°C).

We utilised low-molecular-weight structures of chitin and chitosan (< 3.2 kDa for chitin and < 2.6 kDa for chitosan). Despite the small molecular size, many crucial structural features such as all types of intermolecular interactions identifiable with YASARA software can be observed. At first glance, it is evident that in the case of collagen-chitosan systems, the polysaccharide orientation relative to the peptide helices is much more sensitive to pH than in the case of molecular complexes formed by chitin.

We evaluated the thermodynamic stability of the complexes based on Gibbs free energy of binding. As can be deduced from the negative values of this parameter (Figure 3), the intermolecular complexes are stable under all considered pH conditions. This is consistent with the experimental studies reported by Hua *et al.* [63]. However, in the case of chitosan, its affinity for collagen increases with the basicity of the applied aqueous media, which is indicated by the decreasing relationship between Gibbs free energy of binding and pH characterised by a very good correlation ($R^2 = 0.9$). The highest value corresponding to the smallest mutual affinity of biopolymers is –286.0 kJ/mol, while the lowest value indicating the strongest interactions is –364.8 kJ/mol. The observed trends are understandable considering the acid-base properties of collagen and chitosan. Researchers have documented that chitosan displays typical polyelectrolytic properties that underlie its ability to form complexes with other molecules, such as pharmaceuticals [64] and biopolymers [65, 66]. This specific feature of chitosan is associated with the presence of protonated amino groups (pKa ≈ 6.3 [67]), which enables it to interact with negatively charged species. At a low pH, collagen type II is also positively charged [68]. Therefore, it appears that an increase in the acidic nature of the aqueous medium results in a greater contribution of unfavourable electrostatic repulsion effects between collagen and chitosan. As expected, we noted practically no effect in the case of chitin, which is a non-polyelectrolyte due to the absence of deacetylated amino groups. Furthermore, the significantly higher Gibbs free energy of binding values, ranging from –233.6 to –249.2 kJ/mol, indicate that chitin has a lower affinity for collagen – compared with the affinity of chitosan for collagen – at all pH conditions.

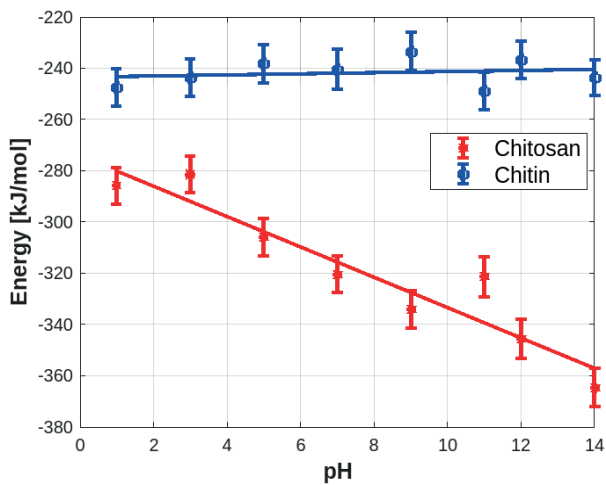

**Figure 3.** The pH effect on the binding free energy calculated for collagen-chitosan and collagen-chitin complexes. The plot also includes error bars, which correspond to the standard deviations.

To describe some structural features related to the molecular complexes we considered in this study, we calculated the number of contacts corresponding to the crucial types of interactions based on the default YASARA structural and energetic settings [53]. When analysing the number of contacts, it is important to note that according to the adopted convention, they express the frequency by which specific interactions occur. For this reason, the number of contacts parameter can be fractional.

According to the YASARA settings, an H–bond is an intermolecular interaction defined by the energy threshold of –6.25 kJ/mol calculated using Equation 2:

$$E_{HB} = -25 \times \frac{2.6 - \max[distance(H-A), 2.1]}{0.5} \times SF_{Donor-Acceptor-H} \times SF_{H-Acceptor-X}, \quad (2)$$

where *SF* and *X* denote a scaling factor and the atom attached to H–bond donor, respectively.

Figure 4 shows the relationship between the number of H–bonds and pH, while Figure 5 shows the distributions of H–bonds formed by different amino acid residues in collagen. As inferred from these data, pH significantly impacts the chitosan-collagen structure, resulting in the creation of a compact lattice of H–bonds under basic conditions. There is no such effect for chitin, a finding consistent with the relationship illustrated in Figure 3. In all cases, the Hyp and Gly moieties form the most H–bonds. This is understandable given that both residues are prevalent in the collagen chain. Gly only plays the acceptor role via the C=O group. We also observed this type of interaction for other residues including Hyp. However, Hyp also forms H–bonds via its –OH group, serving as both a donor and an acceptor.

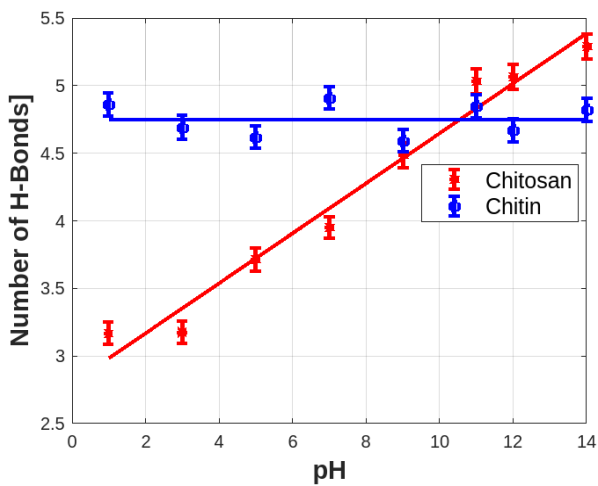

**Figure 4.** The relationship between the number of hydrogen bonds (H-bonds) and pH determined for chitosan and chitin.

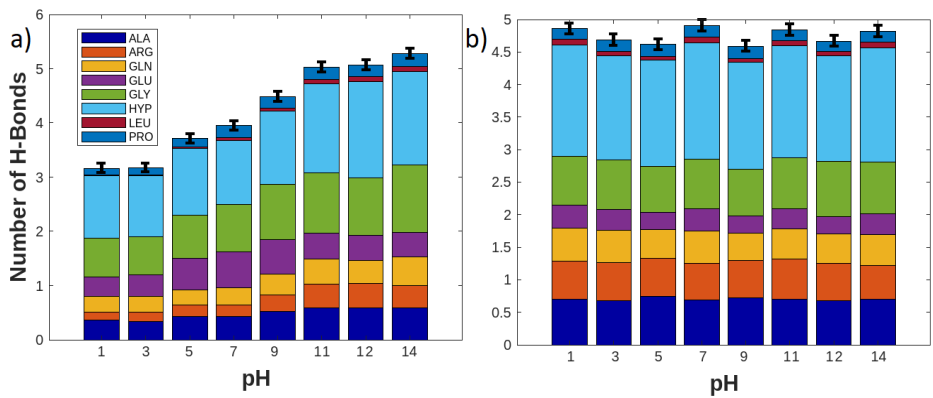

**Figure 5.** The hydrogen bond (H-bond) distributions determined for (a) collagen-chitosan and (b) collagen-chitin complexes.

Another type of interaction we identified is hydrophobic contacts (HP). According to the YASARA definition, these interactions are formed by aliphatic and aromatic carbon atoms. Figure 6 illustrates the correlation between the number of HP interactions and pH, while Figure 7 displays the distributions of HP contacts formed by specific amino acid moieties in collagen. As can be observed, pH has practically no impact on the formation of HP contacts in the case of collagen-chitin complexes. Nevertheless, the total number of such interactions is markedly higher compared with chitosan-collagen complexes. Chitin is widely recognised as a highly hydrophobic biopolymer [69], which aligns with the observed behaviour. The increasing linear trend ($R^2 = 0.97$) in the case of chitosan oligosaccharides can be ascribed to the deprotonation of positively-charged ammonium groups. These effectively solvated moieties are inherently more hydrophilic than non-charged amino groups.

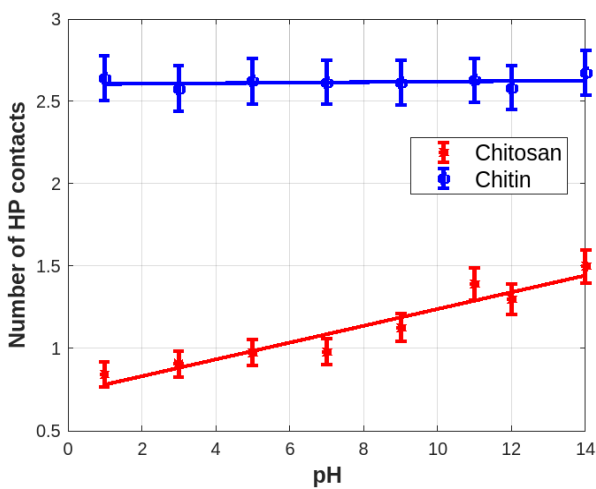

**Figure 6.** The effect of pH on the number of contacts corresponding to hydrophobic interactions in the chitosan-collagen and chitin-collagen complexes.

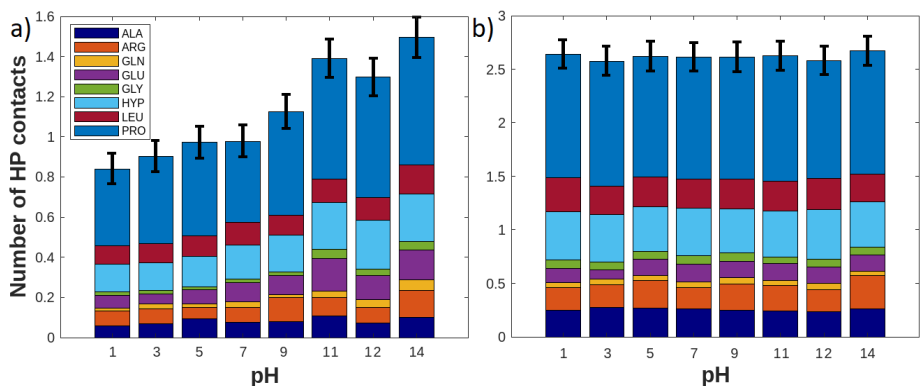

**Figure 7.** The distribution of hydrophobic (HP) intermolecular contacts for the (a) collagen-chitosan and (b) collagen-chitin complexes.

We obtained interesting results regarding the polyelectrolytic properties of chitosan and collagen in the case of ionic interactions. Unlike the other types of molecular forces considered in this study, ionic interactions are not linearly correlated with pH (Figure 8). These interactions are formed between positively charged ammonium groups and negatively charged collagen (ionised carboxylic groups in Arg moieties). The formation of these interactions has been confirmed with infrared spectroscopy [36, 70, 71].

The number of ionic contacts is the highest for pH 5 and 7. This non-linear relationship can be attributed to the fact that proteins are often complex polyelectrolytic systems that are strongly affected by structural features. Of note, the isoelectric point is different for each type of collagen; in general, it ranges from 5 to 9 [72–75]. This indicates that some types of collagen can be negatively charged even when the pH is higher than 5. Of course, in the case of an alkaline environment, chitosan is deprotonated, which makes it difficult to form effective electrostatic interactions. Therefore, there is a specific pH range suitable for ionic contacts.

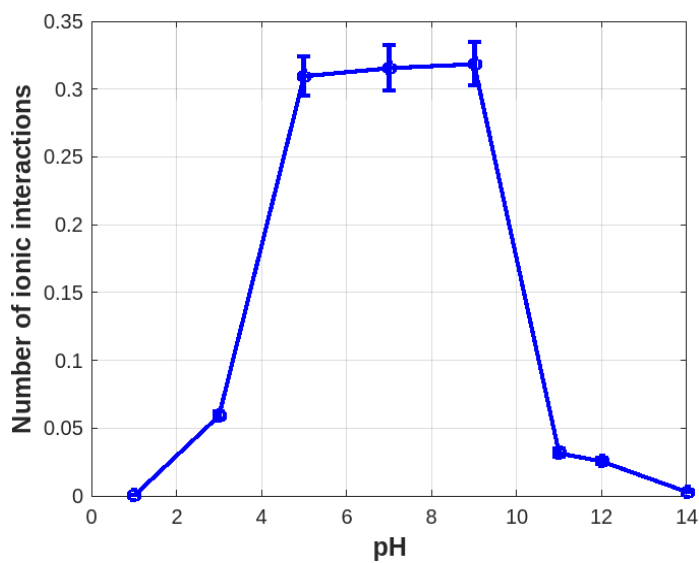

**Figure 8.** The number of ionic intermolecular contacts between collagen and chitosan calculated for different pH.

We next examined the effect of $Na^+$ and $Ca^{2+}$ on the stability of the collagen-chitin and collagen-chitosan complexes. It is well known that multivalent cations including $Ca^{2+}$ can be bound by chitosan, a principle that is applied in wastewater treatment technologies [76–79]. However, the effect of the presence of metal cations on the stability of chitin-collagen and chitosan-collagen complexes has not been described in the literature before. As one can see from Figure 9, the intermolecular interactions formed in chitosan-collagen and chitin-collagen complexes appear to be practically unaffected by $Na^+$ and $Ca^{2+}$. This is understandable considering the very strong interactions discussed above that occur between collagen and chitosan or chitin. On the other hand, there are structurally more complex proteins, such as albumin, whose affinity for polysaccharides relies on the presence of mono- and divalent cations [3].

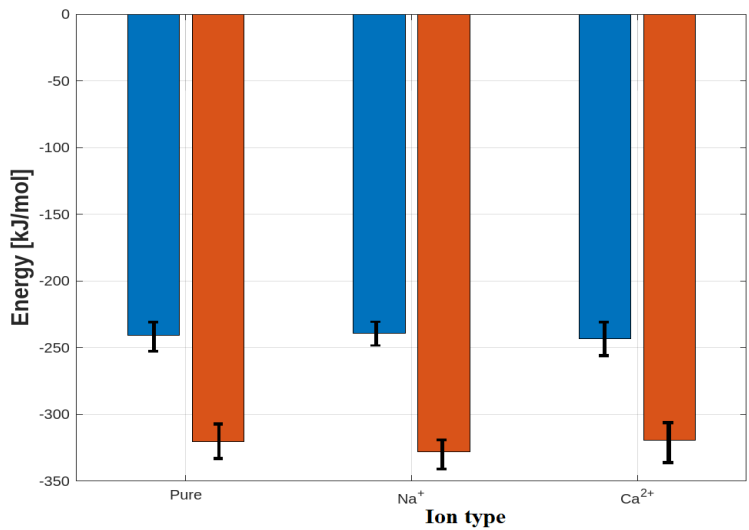

**Figure 9.** The pH effect on the binding free energy calculated for chitosan-collagen and chitin-collagen complexes. The plot also includes error bars, which correspond to the standard deviations.

## 4. Conclusions

We investigated the influence of pH, as well as the presence of $Na^+$ and $Ca^{2+}$, on the affinity of chitosan and chitin oligosaccharides for collagen type II. Furthermore, we analysed some structural features (i.e., the distributions of different types of contacts). We found that pH has a negligible effect on the interactions between collagen and chitin, whose amino groups are all acetylated. On the other hand, we found that the affinity of chitosan for collagen is very sensitive to pH. However, it should be noted that we analysed oligomeric structures. Hence, it is worth investigating in the future the effects of pH and the presence of mono- and divalent cations on intermolecular interactions in large macromolecular assemblies using dedicated coarse-grained methods. Nevertheless, the conventional MD computations seem to be reasonable and consistent with the polyelectrolytic nature of chitosan, contrasting with the absence of such a nature in the case of chitin. The MD simulations also revealed that $Na^+$ and $Ca^{2+}$ have a negligible impact on the affinity between collagen and the studied chitooligosaccharides. To the best of our knowledge, the impact of mono- and divalent cations on the affinity of chitosan and chitin for collagen has not been experimentally investigated. Thus, our results seem to be a good starting point for considering such studies.